# Effect of Y substitution on the structural and magnetic properties of $Dy_{1-x}Y_xCo_5$ compounds

Debjani Banerjee, K. G. Suresh*

Department of Physics, IIT Bombay, Mumbai 400076 India

A. K. Nigam

Tata Institute of Fundamental Research, Homi Bhabha Road, Mumbai 400005, India

## Abstract

Structural and magnetization studies were carried out on $Dy_{1-x}Y_xCo_5$ [x = 0, 0.2, 0.4, 0.6, 0.8, 1] compounds which crystallize in the hexagonal $CaCu_5$-type structure. Lattice parameters and unit-cell volume increase with Y concentration. Large thermomagnetic irreversibility between the field-cooled and the zero-field cooled magnetization data has been observed in all the compounds, which has been attributed to the domain wall pinning effect. Temperature dependence of magnetization data shows that except $DyCo_5$ and $YCo_5$, all the compounds show spin reorientation transitions in the range of 5-300 K. The spin reorientation temperature decreases from 266 K for x=0.2 to 100 K for x=0.8. Powder x-ray diffractograms of the magnetically aligned samples show that $DyCo_5$ has planar anisotropy at room temperature whereas all the other compounds possess axial anisotropy. The spin reorientation transition has been attributed to a change in the easy magnetization direction from the *ab*-plane to the *c*-axis, as the temperature is increased. The anisotropy field and the first order anisotropy constant are found to be quite high in all the compounds except $DyCo_5$. The magnetic properties have been explained by taking into account the variations in contributions arising from the rare earth and transition metal sublattices.

----------------------------------------------------------------

* Corresponding author (email: suresh@phy.iitb.ac.in)

## I. INTRODUCTION

The remarkable hard magnetic properties of $RCo_5$ (R=rare earth) compounds have made them attractive candidates for permanent magnet applications.[1-4] $SmCo_5$ belonging to this series has been identified as one of the superior 'high energy density-high temperature' permanent magnet materials. In an effort to search for materials which do not contain Sm, other $RCo_5$ systems are still being investigated. In this context, structural and magnetic properties of substituted $RCo_{5-x}M_x$ (M= non-magnetic element) and $(R_{1-x}R'_x)Co_5$ phases have been studied.[5-8] Many studies have been reported on $RCo_{5-x}M_x$ compounds with the purpose of revealing the influences of the substitution of M (M=Al, Ga etc.) for Co on both the crystal and magnetic structures[5]. The results indicate that these non-magnetic atoms play an important role in determining the Curie temperatures ($T_C$) and the easy magnetization direction (EMD) of these compounds. It has also been observed that partial substitution of R by another magnetic rare earth having opposite second order Stevens factor ($\alpha_J$) is useful to gain further insight into the fundamental aspects of the magnetism in these compounds. These substitutions sometimes result in interesting phenomena like spin reorientation transitions (SRT), due to changes in their magnetic phase diagram. In this paper, we report the effect of Y substitution on the structure and magnetic properties of the $Dy_{1-x}Y_xCo_5$ compounds. The motivation for choosing Y is to find out the possibility of diluting the planar anisotropy of Dy sublattice so that it can compete with the axial anisotropy of Co sublattice in such a way that the net anisotropy becomes axial at room temperature (RT).

## II. EXPERIMENTAL DETAILS

All the compounds [x = 0, 0.2, 0.4, 0.6, 0.8, 1] were prepared by arc melting the constituent elements of at least 99.9% purity in argon atmosphere. The ingots were melted several times to ensure homogeneity. Subsequently, they were annealed in high purity argon atmosphere at 900 °C for a week. Lattice parameters were determined from the powder x-ray diffraction (XRD) patterns taken using Cu-K alpha radiation. Samples for magnetically aligned XRD measurements were prepared by mixing the powder with an epoxy resin and allowing it to harden in a magnetic field (H) of 10 kOe, applied parallel to the surface of the sample holder. Powder x- ray diffraction patterns of magnetically aligned samples were used to determine the EMD of the compounds. Magnetization (M) measurements were carried out using a vibrating sample magnetometer / SQUID magnetometer in the temperature (T) range 5-300 K, upto a maximum field of 60 kOe. Saturation magnetization ($M_S$) was obtained by plotting M against 1/H in the high field part of the magnetization curves and extrapolating the plot to the 1/H axis. Thermo magnetic analysis was performed from 5-300 K, both under zero-field cooled (ZFC) and field-cooled (FC) modes. In the former case, the samples were cooled in the absence of a field and the magnetization was measured during warming by applying a nominal field of 500 Oe. In the FC mode, the sample was cooled in presence of a field and the magnetization was measured during warming, under the same field. The anisotropy field was estimated by measuring the magnetization on magnetically oriented cylindrical samples. The samples were oriented in a field of 10 kOe and the magnetization was measured by applying the field in directions parallel and perpendicular to the aligning field.

## III. RESULTS AND DISCUSSION

Powder x-ray diffraction patterns show that all the compounds are single phase and crystallize in the hexagonal $CaCu_5$ – type structure. The lattice parameters and the unit cell volume calculated using the Rietveld refinement of the XRD patterns are listed in Table 1. As can be seen, there is an increase in the lattice parameters and the unit cell volume with increase in Y concentration.

Fig. 1a & b show the magnetization isotherms at 5 K and 300 K for all the compounds. It can be seen that the $M_s$ value at 5K is less than that at 300 K in all the compounds except $YCo_5$, which implies that the coupling between the R and TM sublattices is ferrimagnetic in compounds with 0≤x<1 while $YCo_5$ is a ferromagnet. From the M-H plots at 300 K, it can be inferred that the Curie temperatures of this series of compounds is above room

Table 1 Lattice parameters, unit-cell volume, saturation magnetization ($M_S$) and spin reorientation temperature ($T_{SR}$) in $Dy_{1-x}Y_xCo_5$ compounds.

| $x$ | a=b (Å) | c (Å) | V (Å$^3$) | $M_S$(5K) (emu/g) | $M_S$(300K) (emu/g) | $T_{SR}$ (K) |
|---|---|---|---|---|---|---|
| 0 | 4.929 (3) | 3.987 (1) | 83.885 | 30 | 40 | … |
| 0.2 | 4.938 (1) | 3.989 (2) | 84.233 | 34 | 43 | 266 |
| 0.4 | 4.949 (1) | 3.972 (3) | 84.248 | 34 | 51 | 170 |
| 0.6 | 4.952 (2) | 3.970 (1) | 84.308 | 59 | 77 | 166 |
| 0.8 | 4.955 (1) | 3.968 (2) | 84.368 | 72 | 81 | 100 |
| 1 | 4.958 (2) | 3.969 (1) | 84.491 | … | 89 | … |

temperature. It is reported that the $T_C$ of $DyCo_5$ is 973K[9] while that of $YCo_5$ is 977K.[1] It is expected that all the other compounds in this series will have nearly the same value of Curie temperature.

Thermomagnetic plots (Figs 2 a & b) show considerable difference between field cooled and zero field cooled data. A similar difference has been observed in $Er_2Co_{17-x}Al_x$ and $ErCo_{7-x}Cu_x$ compounds recently and can be attributed to domain wall pinning.[10,11] Furthermore the compounds except $DyCo_5$ and $YCo_5$ show anomalies (indicated by arrows) in the M vs. T plots. The temperature corresponding to the anomaly is found to show a systematic decrease as Y concentration increases from 0.2 to 0.8, as can be seen from Table 1. We attribute these anomalies to spin reorientation transition, which arises due to the competition between the Dy and Co sublattice anisotropies. As a general rule, the anisotropy of the rare earth sublattice becomes dominant at low temperatures and may compete with the 3d sublattice anisotropy. Generally the SRT manifest itself as either steps or spikes in the M Vs T curves. In the present case, the SRT temperatures was identified with the inflection points.

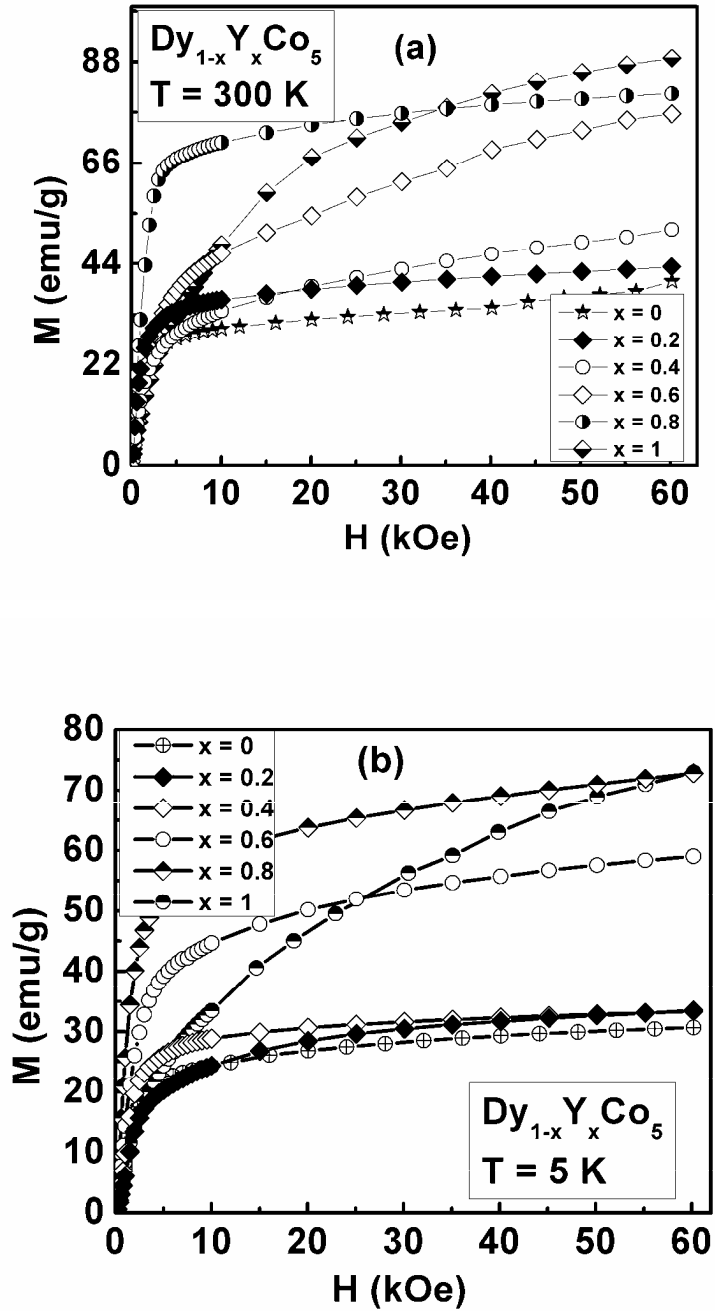

Fig. 1 (a) & (b) M-H isotherms of $Dy_{1-x}Y_xCo_5$ compounds at 300 K & 5 K.

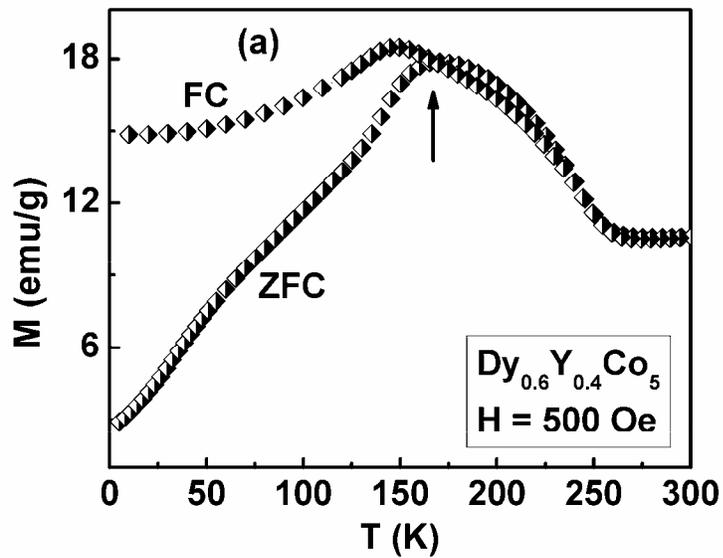

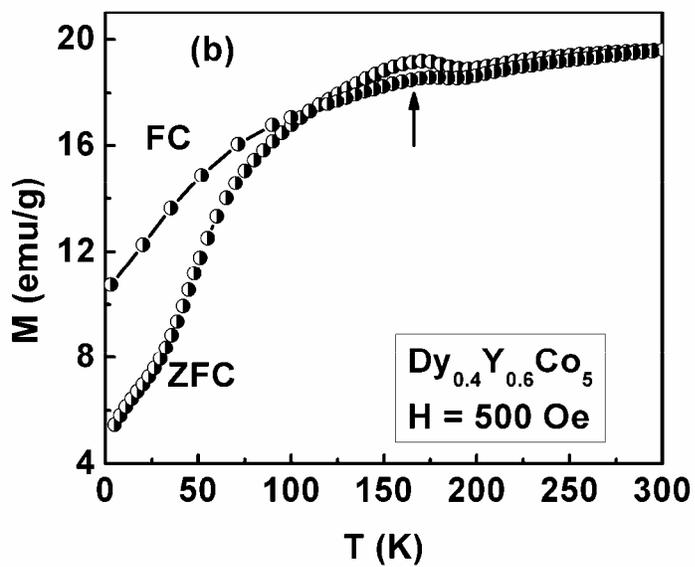

Fig. 2 (a) & (b) M-T plot of $Dy_{1-x}Y_xCo_5$ compound for x = 0.4 & 0.6 both under FC and ZFC conditions. The arrow shows the spin reorientation transition.

In R-TM (transition metal) intermetallic compounds, both the R-sublattice and the T-sublattice contribute to the total anisotropy. The anisotropy of the R-sublattice is induced by the crystalline electric field (CEF) and in the lowest order approximation, it can be expressed as

$$K_1(R) = -\frac{3}{2}\alpha_J \langle r_{4f}^2 \rangle \langle 3J_z^2 - J(J+1) \rangle A_2^0 \quad \ldots\ldots(1)$$

where $\alpha_J$ is the second order Stevens factor, $r_{4f}$ is the mean radius of the 4f electron cloud, J is the total angular momentum and $A_2^0$ is the second-order crystal field parameter. $\alpha_J$ is determined by the degree of asphericity of the 4f electron charge density whereas $A_2^0$ is mainly determined by the charge density around the 4f ion.[12] In general $K_1^R$ is strongly temperature dependent. On the other hand, the anisotropy of the T-sublattice originates from the incompletely quenched angular momentum of the 3d electrons, which is far less temperature dependent well below $T_C$.

In $DyCo_5$, $K_1(Dy)$ is negative, implying planar anisotropy for the Dy sublattice. This is due to the fact that both $A_2^0$ and $\alpha_J$ have negative sign. On the other hand, $K_1(Co)$ is positive for the 1:5 structure[13]. It has been reported that the 3d band dispersion in 1:5 compounds is such that it favours axial anisotropy. This is a consequence of the fact that the axial Co-Co bond lengths are more than the planar bonds. It has been reported that the EMD of $DyCo_5$ changes from planar to axial at 325 K, as the temperature is increased, resulting in SRT[14]. Radwanski has also reported that some of the compounds of $RCo_5$ series with R = Pr, Nd, Tb, Dy and Ho show a rotation of magnetic easy direction from the basal plane to the hexagonal c-axis with increase in temperature.[12] Based on these observations, it may be concluded that the anomalies seen in the present case also correspond to spin reorientation transitions. As can be seen from table 1, the spin reorientation transition temperature ($T_{SR}$) decreases with increase in Y concentration, which is due to the dilution of planar anisotropy of Dy sublattice. Since Y is non-magnetic, its only role here is to reduce the effect of planar contribution from the Dy sublattice. Consequently, the Co sublattice anisotropy starts dominating even at temperatures below the room temperature and hence the $T_{SR}$ decreases with Y concentration. Since the Co sublattice anisotropy is is axial throughout the ferromagnetic temperature region, there is no SRT in the case of $YCo_5$[14].

Extensive studies of spin reorientation and also change of SRT with temperature are also available in pseudobinary $(R_{1-x}R'_x)Co_5$ compounds.[14-16] Tatsumoto et al have reported the spin reorientation of $(Nd_xY_{1-x})Co_5$ compounds.[16] Moze et al. have determined the SRT of pseudobinary $(Pr_{1-x}Nd_x)Co_5$ compounds.[17] Ma et el. have studied the magnetic properties of $(Pr_{1-x}R_x)Co_5$ compounds and have given a series of magnetic phase diagrams.[8] Zhao et al have reported the magnetic phase diagram of all the $RCo_5$ compounds and the role of 3d and 4f sublattice anisotropies in determining the net EMD in each of these compounds.[14] The above reports also confirm the presence of competing magnetocrystalline anisotropies between Co (3d) and R (4f) sublattices.

In order to verify the change in the EMD as a result of spin reorientation transition, we have determined the EMD at room temperature, using the XRD patterns of field-aligned samples. From the aligned XRD patterns shown in Fig. 3, it is observed that for $DyCo_5$, the EMD is in the basal plane as is evident from the retainment of (200) peak and also other (hkl) peaks in the diffractogram of the aligned sample. However, with Y substitution, all peaks except (002) get suppressed, which implies that these compounds have EMD along the c-axis at room temperature. Therefore, aligned XRD study confirms the observations made on the basis of the occurrence of SRT.

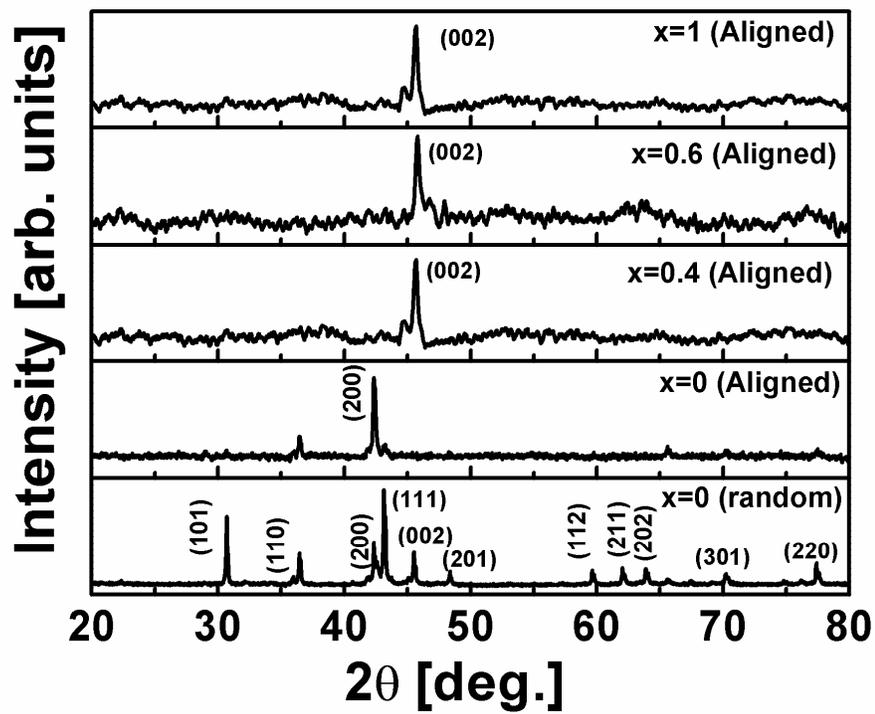

Fig.3. XRD patterns of magnetically aligned samples of $Dy_{1-x}Y_xCo_5$ compounds at RT. The plot at the bottom corresponds to the unaligned sample with x=0.

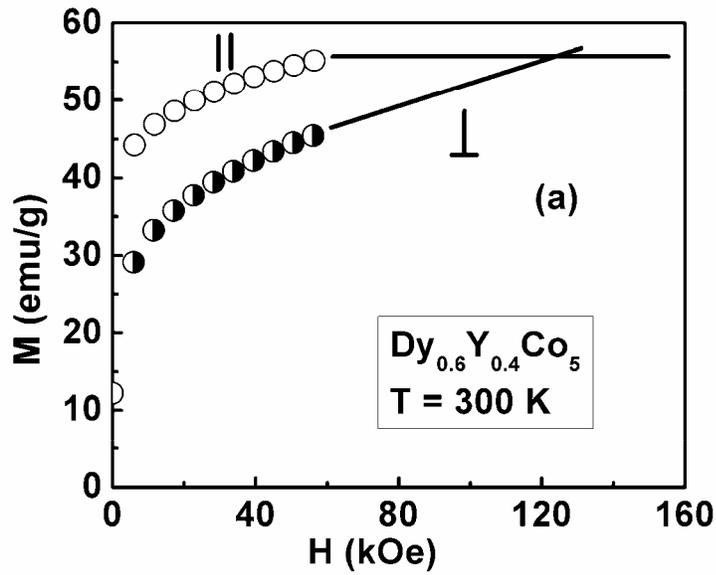
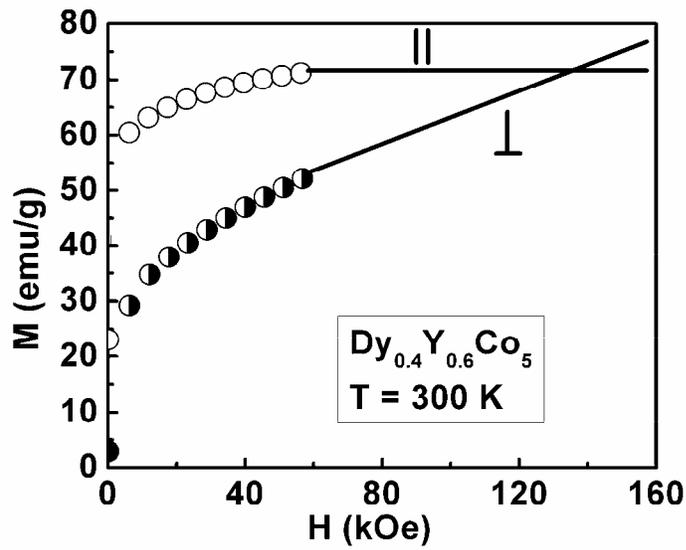

Fig 4 (a) & (b) Magnetization plots of magnetically aligned $Dy_{1-x}Y_xCo_5$ compounds with x=0.4 & 0.6 for the fields applied parallel and perpendicular to the alignment direction, at room temperature.

Table 2. Anisotropy field ($\mu_0 H_A$) and the anisotropy constant ($K_1$) in $Dy_{1-x}Y_xCo_5$ compounds.

| x | $\mu_0 H_A$ (T) [at RT] | $K_1$ (K/f.u.) [at RT] |
|---|---|---|
| 0.4 | 12.2 | 17 |
| 0.6 | 13.5 | 26 |
| 1 | 12.9 | 25 |

The effect of Y substitution on the anisotropy has also been studied by determining the anisotropy field ($H_a$) and the anisotropy constant ($K_1$). This has been done using the magnetization data collected on magnetically aligned samples, with the field applied both parallel and perpendicular to the alignment direction, at RT. The anisotropy field has been determined from the intersection point of the extrapolated magnetization curves. Though the M vs H isotherms were measured upto fields of 60 kOe, the magnetization curves corresponding to the parallel and perpendicular alignments did not intersect each other and hence extrapolation technique was used to calculate $H_a$. Fig. 4 shows the M-H plots of the aligned samples with x = 0.4 & 0.6. Similar plots were obtained for x=1. The anisotropy constant has been calculated using the relation $H_a = 2K_1/M_s$. The variation of $H_a$ and $K_1$ with Y concentration is shown in Table 2. It can be seen that both $H_a$ and $K_1$ increase with Y concentration, which suggests that Y substitution has indeed resulted in an increase in the total axial anisotropy of these compounds. The increase may be due to a reduction in the planar anisotropy of the Dy sublattice mentioend above. The $H_a$ value obtained for $YCo_5$ is in good agreement with the values reported earlier.[4]

## IV. CONCLUSION

In conclusion, all the $Dy_{1-x}Y_xCo_5$ compounds except $DyCo_5$ and $YCo_5$ show anomalies in their M Vs T plots which have been attributed to spin reorientation transitions. The temperature corresponding to the spin reorientation transition decreases with increase in Y concentration because of the dilution of the planar anisotropy of Dy sublattice. Increase of $H_a$ and $K_1$ with Y concentration reflects the enhancement of net axial anisotropy of these compounds.